УДК 528.9

**Т.Е. Самсонов[1], В.Н. Семин[2], П.И.Константинов[3], М.И.Варенцов[4]**
**ВЫЧИСЛЕНИЕ ГЕОМЕТРИЧЕСКИХ ХАРАКТЕРИСТИК ПОДСТИЛАЮЩЕЙ ПОВЕРХНОСТИ И ГОРОДСКОГО КАНЬОНА ДЛЯ МУЛЬТИМАСШТАБНОЙ ПАРАМЕТРИЗАЦИИ МЕТЕОРОЛОГИЧЕСКИХ МОДЕЛЕЙ МЕГАПОЛИСОВ** [5]

**Аннотация**

Представлены результаты исследований по разработке методики вычисления геометрических и тематических характеристик подстилающей поверхности для параметризации локальной модели энерго-массообмена между приземным слоем атмосферы и деятельным слоем подстилающей поверхности урбанизированных территорий (мегаполисов). Получена мультимасштабная база данных параметров подстилающей поверхности с разрешением 200, 500 и 1000 метров. Использование микрометеорологических моделей, учитывающих специфику городской среды в сочетании с мезомасштабными прогностическими моделями позволит заметно улучшить качество воспроизведения метеорологических полей и локального оперативного прогноза погоды в мегаполисах, где особо важен учет гидрометеорологической обстановки.

**Ключевые слова**: городской микроклимат, городской каньон, подстилающая поверхность, пространственный анализ.

[1]Московский государственный университет имени М.В. Ломоносова, географический факультет, кафедра картографии и геоинформатики, лаборатория автоматизации, ст. науч. сотр., Ярославский государственный университет, Лаборатория дискретной и вычислительной геометрии им. Делоне, мл. науч. сотр., канд. геогр. н. *e-mail:* tsamsonov@geogr.msu.ru

[2]Московский государственный университет имени М.В. Ломоносова, географический факультет, кафедра картографии и геоинформатики, лаборатория автоматизации, науч.сотр., *e-mail:* vnsemin@mail.ru

[3]Московский государственный университет имени М.В. Ломоносова, географический факультет, кафедра метеорологии и климатологии, ст.преподаватель, канд. геогр. н. *e-mail:* kostadini@mail.ru

[4]Московский государственный университет имени М.В. Ломоносова, географический факультет, кафедра метеорологии и климатологии, студент *e-mail:* mvar91@gmail.com

# Введение

Городская застройка вносит заметное искажение в направление и силу воздушных потоков в нижней атмосфере. И, если при моделировании потоков тепла и влаги над асфальтом, водой и даже над лесом можно ограничиться более детальным описанием подстилающей поверхности на уровне теплофизических свойств и параметра шероховатости (Oke, 1987), то в пределах модельных ячеек с городской застройкой необходимо прибегать к гораздо более сложным параметризациям, учитывающим геометрию застройки, а также реальное соотношение различных типов подстилающей поверхности (здания, вода, растительность и т.п.).

Городской каньон, согласно теории, разработанной во второй половине XX века канадским ученым Тимом Оке (Nunez, Oke 1977), представляет собой упрощенную геометрическую форму улицы в профиль, имеющей два борта (стены домов) и днище (сама улица). Ключевые параметры такой структуры – соотношение высоты бортов к ширине днища и азимутальная ориентация осевой линии.

Зависимость метеорологического режима каньона от его характеристик не на климатическом, а на оперативном временном масштабе пока слабо изучена. Практически все существующие модели городского пограничного слоя атмосферы (Masson, 2000; Kusaka 2001; Martilli, 2002) учитывают городской каньон как одну из форм подстилающей поверхности, наряду с парками и водными объектами. В некоторых моделях весь город упрощенно рассматривается как определенным образом ориентированный каньон. Однако такой подход хорош для населенных пунктов, которые имеют правильный характер застройки, и где ориентация каньонов одинакова в большей части мегаполиса. Для городов, обладающих более хаотичной структурой, в работе (Masson, 2000) была предложена упрощенная схема расчета теплового баланса, основанная на осреднении используемых формул путем интегрирования по всем возможным азимутам ориентации каньона. Однако для многих мегаполисов, например для Москвы, где преобладает радиальный характер застройки, такое осреднение корректно только для расчета среднесуточных величин и, соответственно, для климатических прогнозов (Кислов, Константинов, 2011). Существует опыт интеграции относительно простых параметризаций городской застройки в мезомасштабные модели, такие как WRF (Skamarock, 2007). С их помощью осуществляется как детализация прогноза погоды и климата на более крупном пространственном масштабе, так и изучение микроклимата улиц и микрорайонов мегаполисов с целью улучшения планировки городов.

В течение последних лет проводились работы по исследованию и моделированию микроклимата внутри городской застройки на примере Московского мегаполиса (Кислов, Константинов, 2011). Территория города была разбита на сетку ячеек с разрешением 500 метров, что позволило учесть пестроту городской подстилающей поверхности. При этом был произведен анализ категорий землепользования и составленная классификация, учитывающая водные поверхности, промзоны, лесопарки, а также плотно застроенные территории. Данная модель позволила сделать климатический прогноз термического режима Москвы на XXI век.

Для оперативных прогнозов погоды, для которых важно воспроизведение суточного хода метеовеличин, для Москвы необходим более детальный учет характеристик подстилающей поверхности и ориентации каньона, определяющий время и интенсивность освещения солнцем его поверхностей, что требует привлечения данных ГИС о реальной застройке мегаполиса. Подобные данные, получаемые по результатам анализа цифровых карт и данных дистанционного зондирования, играют ведущую роль в параметризации метеорологических моделей мегаполисов, наряду с натурными наблюдениями. Исследования на их основе отличаются многогранностью аспектов, которые мы кратко рассмотрим.

В работах (Voogt, Oke, 2003; Weng, 2009) проведен детальный анализ возможностей космической съемки в тепловом диапазоне для исследования городского острова тепла. Найдена зависимость интенсивности городского острова тепла от соотношения высоты «среднего» каньона к его ширине. Показаны сложности в развитии методов, связанные с использованием в моделировании численных характеристик землепользования вместо более фундаментальных физических дескрипторов поверхности.

Балдина и Грищенко (2011) на основе тепловых снимков Landsat ETM+ исследовали тепловой остров Москвы, выявив при этом особенности его отображения на разносезонных снимках, такие как бимодальное распределение яркостей в теплый период года, зависимость кривой распределения от годового хода температуры. Кроме этого, им удалось построить временные образы тепловых аномалий различных объектов, таких как водные объекты, промышленные зоны, лесные и естественные открытые поверхности, жилая застройка. Эти образы позволяют на каждый сезон года получать распределение интенсивности излучения между городскими объектами.

Влияние характера подстилающей поверхности на свойства тепловых потоков в городской среде оценено в работе (Hagishima, Tanimoto, 2005) с целью формирования модели городского полога (urban canopy model), необходимой для параметризации метеорологических моделей в урбанизированных территориях. Подобная параметризация использована в работе (Kondo et al., 2008) для вычисления потоков ветра в мезомасштабной модели AIST-MM на примера города Токио. Данные об открытости небосвода и высоте зданий, полученные по результатам анализа цифровых карт Токио были осреднены на расчетную сетку.

В работе (Jackson et al., 2010) описаны методы и характеристики, использованные при создании набора данных, который может быть использован для оценки влияния урбанизированных территорий в глобальных климатических моделях. В числе характеристик оказались городская морфология, радиационные характеристики зданий и прочие параметры. Плотность застройки определялась по данным дистанционного зондирования. Исследования показали увеличение роли островов тепла в высоких широтах.

Моделирование аэродинамических параметров массивов зданий при различных их конфигурациях проведено в работе (Zaki et al., 2011). При этом случайным образом задавалась высота и поворот зданий. Влияние вертикальной и горизонтальной случайности распределения зданий на аэродинамические параметры объяснено структурой появляющихся вихрей в кварталах. Использование данных дистанционного зондирования в работе (Chow et al., 2011) позволило создать трехмерную модель подстилающей поверхности для моделирования эффекта локального острова холода, создаваемого небольшим парком в городе аридной зоны.

Сочетание статистических показателей по городским районам и физических характеристик, извлеченных из данных дистанционного зондирования на город Индианаполис позволило исследователям в работе (Liang, Weng, 2011) построить синтетический индикатор, отражающий изменения качества городской среды за последнее десятилетие. Цифровые карты разных лет, а также свежие ортофотоснимки города Лиссабон позволили оценить рост плотности застройки с 1960 г. по 2004 методом пространственно-временного фрактального анализа (Encarnacao et al., 2012). Результаты предназначаются прежде всего для осуществления градостроительного планирования и регулирования.

Таким образом, можно видеть необычайное разнообразие работ, посвященных изучению влияния городов на климат с привлечением пространственных данных. В то же время, современная городская метеорология требует наполнения баз данных по характеристикам подстилающей поверхности крупных городов. Одновременно с этим, расчет параметров городского каньона в среде ГИС требует разработки специализированных методов и инструментов пространственного анализа.

В данном исследовании мы уделяем внимание расчетам детальных характеристик подстилающей поверхности и городского каньона на основе картографических данных, а также созданию базы данных, включающей эти характеристики на территорию города Москвы. Для их вычисления использованы такие средства пространственного анализа как оверлей, триангуляция и цифровое моделирование рельефа.

**Характеристики подстилающей поверхности и городского каньона**

Перечень необходимых параметров определяется требованиями входных данных для метеорологического моделирования. В нашем случае вычисления проводились для параметризации разрабатываемых моделей экстремальных скоростей ветра в городских каньонах, а также энерго-массообмена между приземным слоем атмосферы и подстилающей поверхностью. Список характеристик представлен в **Таблице 1**, которая отражает схему таблиц, получаемых по результатам расчетов. Характеристики 2–13 относятся к подстилающей поверхности, 14–35 описывают геометрию городского каньона, 36–42 хранят вспомогательную информацию. Идентификатор хранится в поле 1.

**Таблица 1. Параметры подстилающей поверхности**

| № | Название | Описание | Ед. изм. |
|---|---|---|---|
| 1 | OBJECTID | Идентификатор | 0—... |
| 2 | BLD_AREA | Площадь зданий | м² |
| 3 | BLD_RATIO | Доля зданий в площади ячейки | [0, 1] |
| 4 | GREEN_AREA | Площадь озелененных территорий | м² |
| 5 | GREEN_RATIO | Доля озелененных территорий в площади ячейки | [0, 1] |
| 6 | INDUSTR_AREA | Площадь промышленных территорий и площадей | м² |
| 7 | INDUSTR_RATIO | Доля промышленных территорий и площадей в площади ячейки | [0, 1] |
| 8 | ROAD_AREA | Площадь дорог | м² |
| 9 | ROAD_RATIO | Доля дорог в площади ячейки | [0, 1] |
| 10 | WATER_AREA | Площадь водных объектов | м² |
| 11 | WATER_RATIO | Доля водных объектов в площади ячейки | [0, 1] |
| 12 | OTHER_AREA | Площадь прочих объектов | м² |
| 13 | OTHER_RATIO | Доля прочих объектов в площади ячейки | [0, 1] |
| 14 | SVF_MEAN | Среднее значение открытости небосвода | [0, 1] |
| 15 | SVF_NOBLD_MEAN | Среднее значение открытости небосвода без учета крыш зданий | [0, 1] |
| 16 | BLD_MEAN_HEIGHT | Средневзвешенная высота здания | м |
| 17 | MDC_WIDTH | Средняя ширина направленного каньона | м |
| 18 | MDC_AREA | Площадь направленных каньонов | м² |
| 19 | MDC_RATIO | Доля направленных каньонов в общей площади каньонов | [0, 1] |
| 20 | MUC_WIDTH | Средняя ширина ненаправленных каньонов | м |
| 21 | MUC_AREA | Площадь ненаправленных каньонов | м² |
| 22 | MUC_RATIO | Доля ненаправленных каньонов в общей площади каньонов | [0, 1] |
| 23 | BLDC_RATIO | Доля зданий, образующих направленные каньоны | [0, 1] |
| 24 | BLDUC_RATIO | Доля зданий, образующих направленные и ненаправленные каньоны | [0, 1] |
| 25 | BLUC_RATIO | Доля зданий, образующих ненаправленные каньоны | [0, 1] |
| 26 | FRONT_INDEX | Фронтальный индекс для направленных каньонов (отношение длин стен и просветов между ними) | [0, 1] |
| 27 | DIR1_6 | Первое преобладающее направление при расчете по 6 секторам (30°) | 0° — 180° |
| 28 | DIR2_6 | Второе преобладающее направление при расчете по 6 секторам (30°) | 0° до 180° |
| 29 | DIRRATIO_6 | Отношение повторяемости первого и второго преобладающих направлений при расчете по 6 секторам (30°) | 1—... |
| 30 | DIR1_7 | Первое преобладающее направление при расчете по 7 секторам (25,7°) | 0° — 180° |
| 31 | DIR2_7 | Второе преобладающее направление при расчете по 7 секторам (25,7°) | 0° — 180° |
| 32 | DIRRATIO_7 | Отношение повторяемости первого и второго преобладающих направлений при расчете по 7 секторам (25,7°) | 1—... |

| 33 | DIR1_8 | Первое преобладающее направление при расчете по 8 секторам (22,5°) | 0° — 180° |
| 34 | DIR2_8 | Второе преобладающее направление при расчете по 8 секторам (22,5°) | 0° — 180° |
| 35 | DIRRATIO_8 | Отношение повторяемости первого и второго преобладающих направлений при расчете по 8 секторам (22,5°) | 1—… |
| 36 | X | Прямоугольная координата X центра ячейки в проекции UTM N37 | м |
| 37 | Y | Прямоугольная координата Y центра ячейки в проекции UTM N37 | м |
| 38 | LAT | Геодезическая широта центра ячейки на эллипсоиде WGS84 | ° |
| 39 | LONG | Геодезическая долгота центра ячейки на эллипсоиде WGS84 | ° |
| 40 | Z_MEAN | Средняя абсолютная отметка ячейки над уровнем моря | м |
| 41 | SHAPE_Length | Периметр ячейки | м |
| 42 | SHAPE_Area | Площадь ячейки | м$^2$ |

Структура таблиц одинакова для сеток разных разрешений — 1000, 500 и 200 м. Последовательное уменьшение размера ячейки позволяет осуществлять даунскейлинг мезометеорологических моделей типа COSMO и WRF на микрометеорологический масштаб и уточнять прогноз погоды локально. Таким образом, получаемые нами данные предназначены для мультимасштабной параметризации метеорологических моделей.

В качестве источников данных для вычисления параметров мы использовали карты OpenStreetMap, а также цифровые модели рельефа SRTM90 и ASTER GDEM v.2.

Решение задач производилось средствами ГИС-пакета *ArcGIS for Desktop 10.1*, путем разработки дополнительных модулей средствами *ArcGIS SDK* на языке *C++*, а также отдельных программных решений на основе библиотеки *GeoTools* и языка программирования *Java*.

**Вычисление характеристик подстилающей поверхности**

Различные типы подстилающей поверхности (параметры 2—13) обладают неодинаковой влагопроницаемостью и отражательной способностью (альбедо), и, следовательно определяют локальные особенности энергомасообмена с приземным слоем атмосферы. Для параметризации метеорологических моделей городских территорий на микро- и мезомасштабе важны следующие типы подстилающей поверхности: здания (крыши домов), лесная растительность, твердое покрытие (асфальт), естественное покрытие (включая луговую растительность), водные объекты.

Четкое разграничение твердого и естественного покрытия при использовании картографических данных возможно только на участках, занимаемых городскими улицами. Заасфальтированные площадки в пределах промышленных территорий могут быть выделены с привлечением данных дистанционного зондирования. Поскольку это отдельная задача, выходящая за рамки исследования, мы выделили особый тип подстилающей

поверхности — промышленные территории, приняв для него соотношение естественного и твердого покрытия 1:1. К этой же категории были отнесены городские площади, для которых также не ясно распределение твердого и естественного покрытия.

Древесная растительность и естественное покрытие не всегда однозначно дифференцируются по картографическим данным. Классификаторы, присутствующие в используемых нами данных, содержат множество типов озелененных территорий — парки, лесопарки, леса, парки культуры и отдыха, скверы, бульвары, леса — некоторые из которых могут содержать в своих пределах участки как древесной растительности, так и естественного покрытия в виде открытого грунта, полян и т.д. Мы условно приняли все озелененные территории как территории древесной растительности, что оправдано, так как влияние на метеорологический режим оказывают крупные массивы растительности, соответствующие лесопаркам и лесам, отнесение которых к типу древесной растительности сомнений не вызывает.

Типы подстилающей поверхности зданий и водных объектов выделяются по картографическим данным однозначно. Все остальные территории в нашем покрытии были отнесены к прочим и приняты за естественное покрытие. Пример классификации территории по типам подстилающей поверхности приведен на **Рис. 1**.

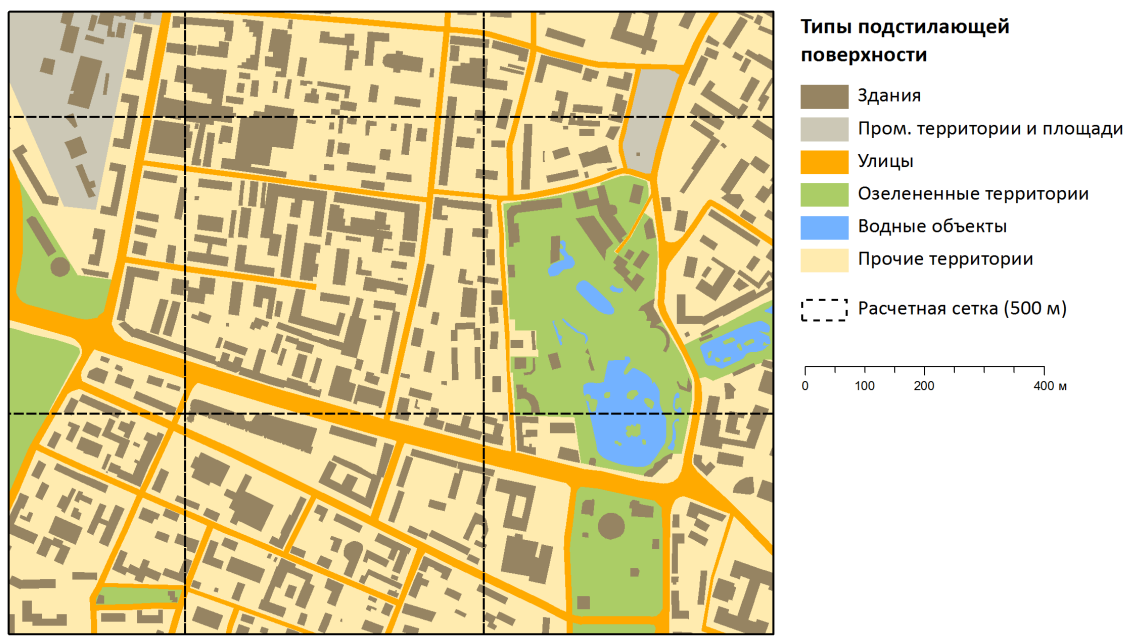

**Рис. 1. Типы подстилающей поверхности**

После того, как определен классификатор типов, для каждой модельной ячейки необходимо определить соотношение площадей каждого из них. Для этого была разработана следующая последовательность действий:

1. Интеграция зданий в слой топографического районирования посредством оверлея в режиме объединения (*Union*).
2. Реклассификация картографических данных в соответствии с искомыми типами подстилающей поверхности (*Reclass*).

3. Формирование расчетных сеток с регулярным шагом на нескольких уровнях детализации — 200, 500, 1000 м.
4. Оверлей каждой расчетной сетки и слоя топографического районирования в режиме пересечения (*Intersect*).
5. Вычисление отношения площади каждого типа подстилающей поверхности и площади ячейки для каждого уровня детализации.
6. Присоединение полученных площадей и долей типов подстилающей поверхности к слою расчетной сетки.

По результатам расчетов в базе данных для каждой ячейки наполняются поля, содержащие площади различных типов подстилающей поверхности и их доля в площади ячейки.

На Рис. 2 приведена картограмма доли зданий в площади ячейки для сетки с шагом 500 метров, построенная по результатам расчетов.

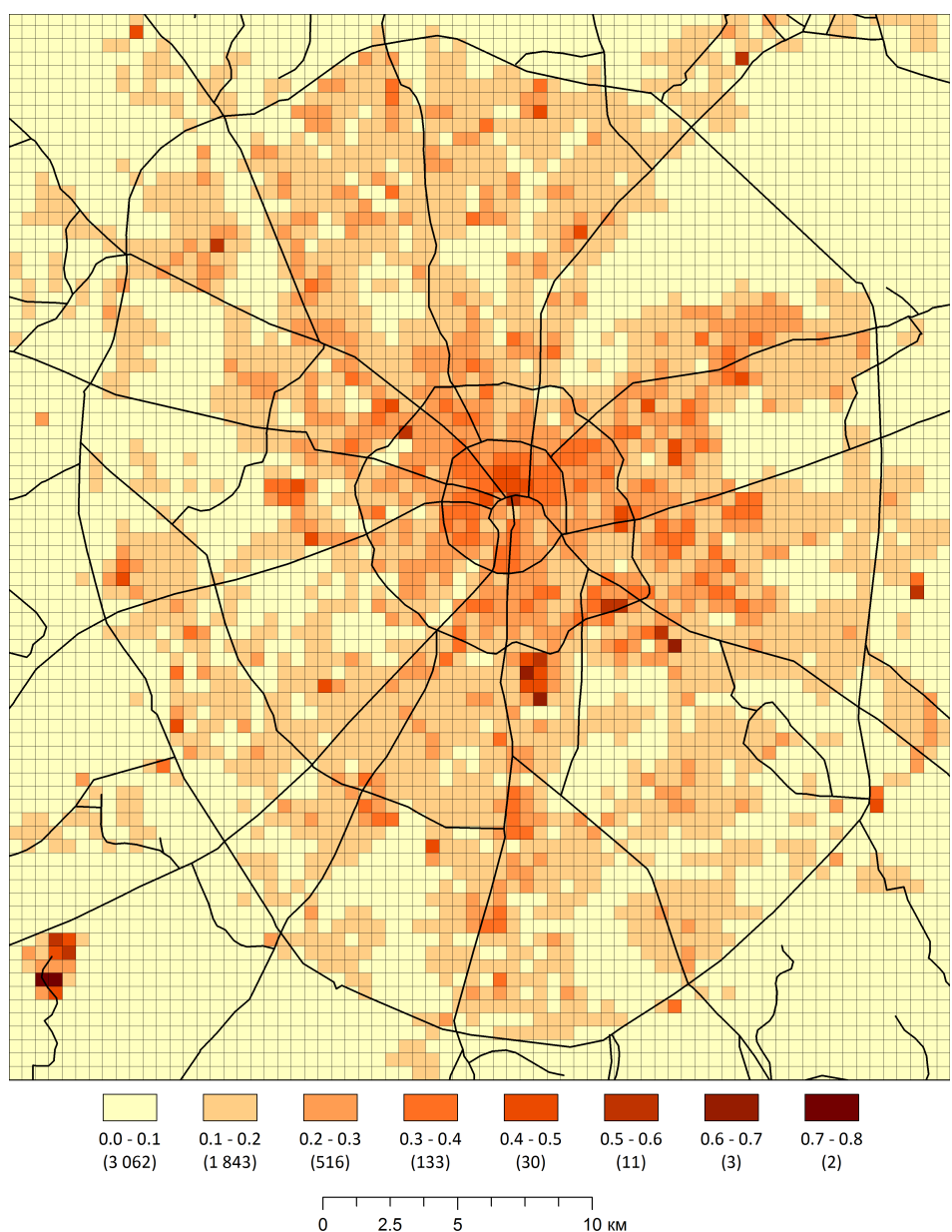

**Рис. 2. Доля зданий в площади ячейки. В скобках указано число ячеек каждого класса. Разрешение сетки 500 м.**

**Вычисление характеристик городского каньона**

В отличие от параметров подстилающей поверхности, которые могут быть определены напрямую по данным дистанционного зондирования, вычисление геометрических характеристик городского каньона (14-35, Таблица 1) затруднительно без использования картографических данных. Детальность описания каньона обусловлена необходимостью статистического учета аэродинамических и тепловых эффектов, вызываемых городской застройкой, таких как усиление ветра и перераспределение прямой солнечной радиации. Все вычисляемые параметры можно поделить на 4 группы: открытость небосвода, размеры и пропорции каньонов, функциональная роль зданий в каньонах и преобладающие направления каньонов.

*Открытость небосвода (параметры 14–15)*

Открытость небосвода — это доля небесной полусферы, видимая из точки. Для обозначения открытости небосвода обычно применяется англоязычная аббревиатура SVF (sky view factor). Для аппроксимации этой величины применяются различные расчетные схемы. Наиболее простой подход заключается в вычислении среднего значения косинуса угла по *n* направлениям с равным шагом:

$$\Psi = \frac{1}{n}\sum_{i=1}^{n}\cos v_i,$$

где *n* — число направлений, на которое разбит круг 360 градусов. При неравномерности шага расчетные формулы усложняются.

Расчет SVF в нашей методике осуществляется на основе растровой модели данных. Полигональный слой зданий преобразуется в растр, значения которого в ячейках соответствуют высоте дома. Полученный растр складывается с цифровой моделью рельефа, что позволяет внедрить здания в качестве искусственных неровностей рельефа. Чтобы уловить каньоны, обычно достаточно разрешения 5 метров. Более грубое разрешение (10 и более метров) может привести к слиянию зданий на плотно застроенных территориях, где преобладают гаражные кооперативы, заводские территории, узкие переулки.

При вычислении открытости небосвода задается два параметра: число секций и радиус поиска точки. В круговом секторе, ограниченном азимутальными направлениями и окружностью заданного радиуса, находится точка (пиксел), образующая максимальный угол затенения. Косинус этого угла дает показатель открытости небосвода по выбранному направлению, а их среднее арифметическое по всем секторам — оценку общей открытости небосвода.

Фрагмент цифровой модели открытости небосвода, полученной данным методом, представлен на Рис. 3.

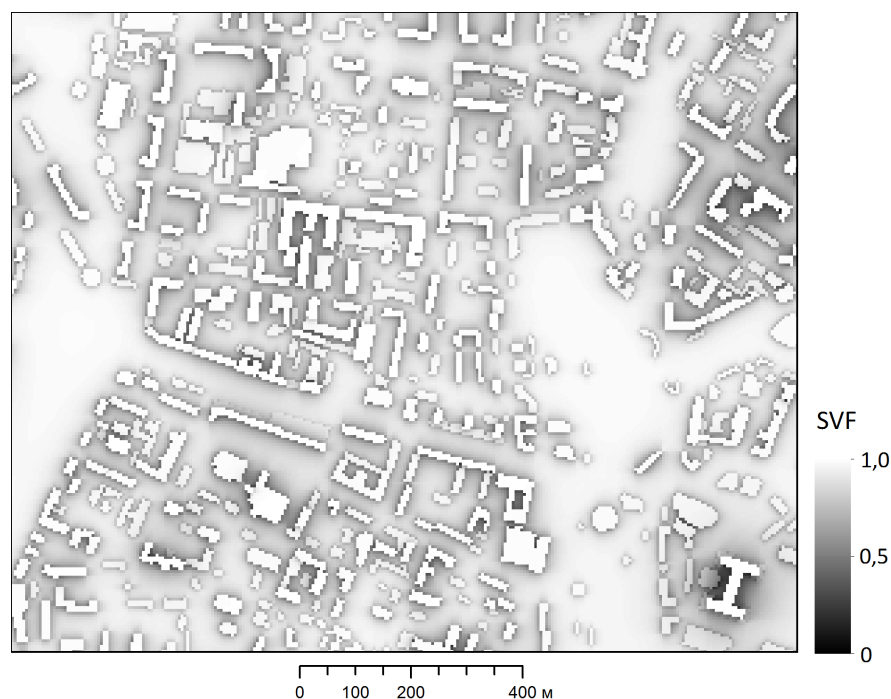

**Рис. 3. Открытость небосвода (SVF)**

*Размеры и пропорции каньонов (параметры 16–22)*

Локальная геометрия городского каньона определяется двумя параметрами: расстоянием между домами и их высотой. Основные характеристики каньона — средняя высота, средняя ширина, площадь. Метеорологическое моделирование требует их отдельного расчета по каньонам двух типов: направленным и ненаправленным.

Для выделения границ каньонов мы использовали триангуляцию Делоне с ограничениями. В качестве исходных точек использовались углы домов, а стены домов и границы лесных массивов использовались в виде жестко зафиксированных ребер. При подходе, основанном на триангуляции, направленные каньоны в первом приближении соответствуют треугольникам, пересекаемым линиями автомобильных дорог и содержащим как минимум один угол здания. Ненаправленные каньоны соответствуют треугольникам, содержащим как минимум один угол здания, но не пересекаемым автомобильными дорогами. Треугольники, не содержащие углов зданий, к каньонам не относятся. Различные типы каньонов, выделенные таким способом, представлены на Рис. 4.

За локальную ширину каньона удобно принять высоту треугольника, соответствующую кратчайшей стороне:

$$w_i = \frac{2S_i}{\min(a_i, b_i, c_i)},$$

где $S_i$ — площадь $i$-го треугольника, $a_i$, $b_i$, $c_i$ — стороны $i$-го треугольника. Тогда средняя ширина каньона в пределах ячейки может быть найдена как:

$$\bar{w} = \frac{1}{n}\sum_{i=1}^{n} w_i$$

где *n* — число треугольников в пределах ячейки. Площадь, занимаемая каньонами, находится путем суммирования $S_i$.

В качестве характеристики, описывающей среднюю высоту каньона мы использовали средневзвешенную по площади высоту зданий:

$$\bar{h} = \sum_{i=1}^{n} s_i h_i \bigg/ \sum_{i=1}^{n} s_i ,$$

где *n* — число зданий внутри ячейки, $s_i$ — площадь *i*-го здания, $h_i$ — высота *i*-го здания. Пропорции каньона определяются отношением средней ширины и высоты.

*Функциональная роль зданий в каньонах (параметры 23–26)*

Здания внутри ячейки формируют каньоны, которые отличаются устойчивостью направления и сомкнутостью стен. С точки зрения устойчивости направления выделяют направленные и ненаправленные каньоны (см. предыдущий параграф). При этом здания могут формировать каньоны либо какого-то определенного типа, либо обоих типов. Для определения соотношения зданий этих групп считается доля площади зданий, образующих направленные каньоны (23), каньоны обоих типов (24) и ненаправленные каньоны (25). Отбор зданий каждого типа осуществляется путем пространственного запроса пересечения с треугольниками направленных и ненаправленных каньонов. Пример функционального разделения зданий по типам формируемых ими каньонов представлен на Рис. 4.

Сомкнутость стен каньонов оценивается соотношением стен домов и просветов между ними по границе каньона. Эту величину в городской метеорологии принято называть *фронтальным индексом*. Для вычисления фронтального индекса отбираются ребра треугольников, идущие по внешней границе направленных каньонов. Далее путем пространственного запроса помечаются те ребра, которые совпадают со стенами зданий. Остальные ребра соединяют углы зданий и заполняют просветы между ними. При таком подходе фронтальный индекс каньонов *F* определяется отношением:

$$F = \sum_{i=1}^{n} l_i \bigg/ \sum_{j=1}^{m} l_j ,$$

где *n* — число ребер, идущих вдоль зданий, *m* — число ребер, идущих между зданиями, $l_i$ — длина i-го ребра, идущего вдоль здания, $l_j$ — длина j-го ребра, идущего между зданиями.

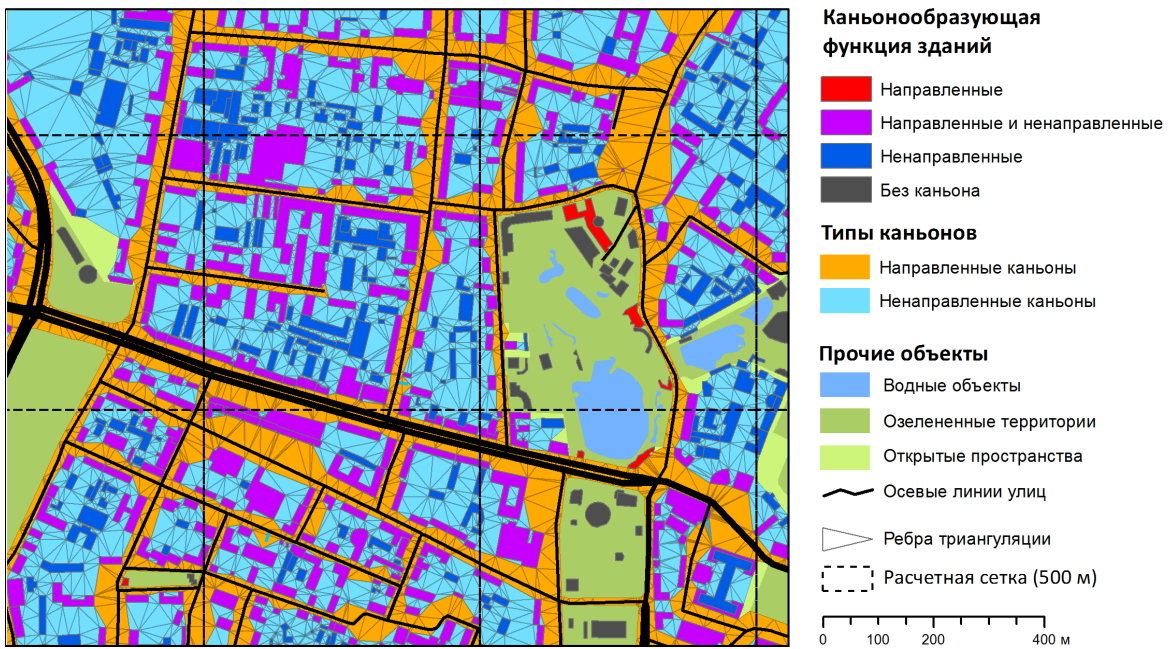

**Рис. 4.** Типы каньонов и каньонообразующих зданий

*Преобладающие направления каньонов (параметры 27–35)*

Преобладающее направление каньона оказывает влияние на перемещение приземных воздушных масс. При этом для городской застройки типично не одно, а два основных направления улиц, которые обычно перпендикулярны. Эти направления соответствуют модам распределения азимутов улиц в пределах ячейки, а отношение частот этих мод характеризует преобладание одного направления над другим. Для вычисления преобладающих направлений и их повторяемости был применен следующий алгоритм:

1. Разрезать линии улиц расчетной сеткой, присвоить каждой линии идентификатор квадрата (оверлей в режиме *Identity*).
2. Разбить каждую линию в узлах на элементарные отрезки с двумя вершинами.
3. Установить число классов $N$, по которым будет строиться гистограмма направлений улиц, и соответствующую ему величину интервала $h$.
4. Для каждого $i$-го класса, ограниченного значениями азимутов $a_i$ и $a_{i+1}$ определить сумму длин отрезков $L_i$, попадающих в данный диапазон:

$$L_i = \sum_j l_j : a_j \in [a_i, a_{i+1})$$

5. Найти номер класса $i = m$ в котором величина $L_i$ максимальна. Принять частоту первой моды равной $L_m$.
6. Найти номер класса $r > k$, образующий пик гистограммы со значением $L_r$ справа от $L_m$. Принять частоту второй моды равной $L_r$
7. Найти номер класса $l < k$, образующий пик гистограммы со значением $L_l$ слева от $L_m$. Если $L_l > L_r$, принять частоту второй моды равной $L_l$.
8. Вычислить значение первой и второй моды по формуле:

$$M_i = a_i + h \frac{L_i - L_{i-1}}{2L_i - L_{i-1} - L_{i+1}}, \text{ для } i = m, r\ (l).$$

9. Вычислить отношение частот первой и второй моды $f_M = M_m / M_{l(r)}$.

Известно, что значение моды зависит от числа и границ интервалов, в которых подсчитывается количество входящих объектов. Чтобы снизить влияние этого эффекта, расчеты производятся при различных значениях интервалов. В нашем случае использовалось разбиение полукруга на $N = 6$, 7 и 8 секторов, что соответствует величинам интервалов $h = 30°$, 25,7° и 22,5°. По результатам расчетов в соответствующие поля записываются значения $M_m$, $M_{l(r)}$ и $f_M$, а при введении в модель в качестве параметров используются их осредненные значения.

Фрагмент карты рассчитанных значений преобладающих направлений каньонов представлен на Рис. 5. В некоторых ячейках отсутствует второе преобладающее направление. Это означает, что на гистограмме азимутов присутствует только один положительный экстремум. Также есть несколько ячеек, в которых отсутствуют каньоны и направление не определено.

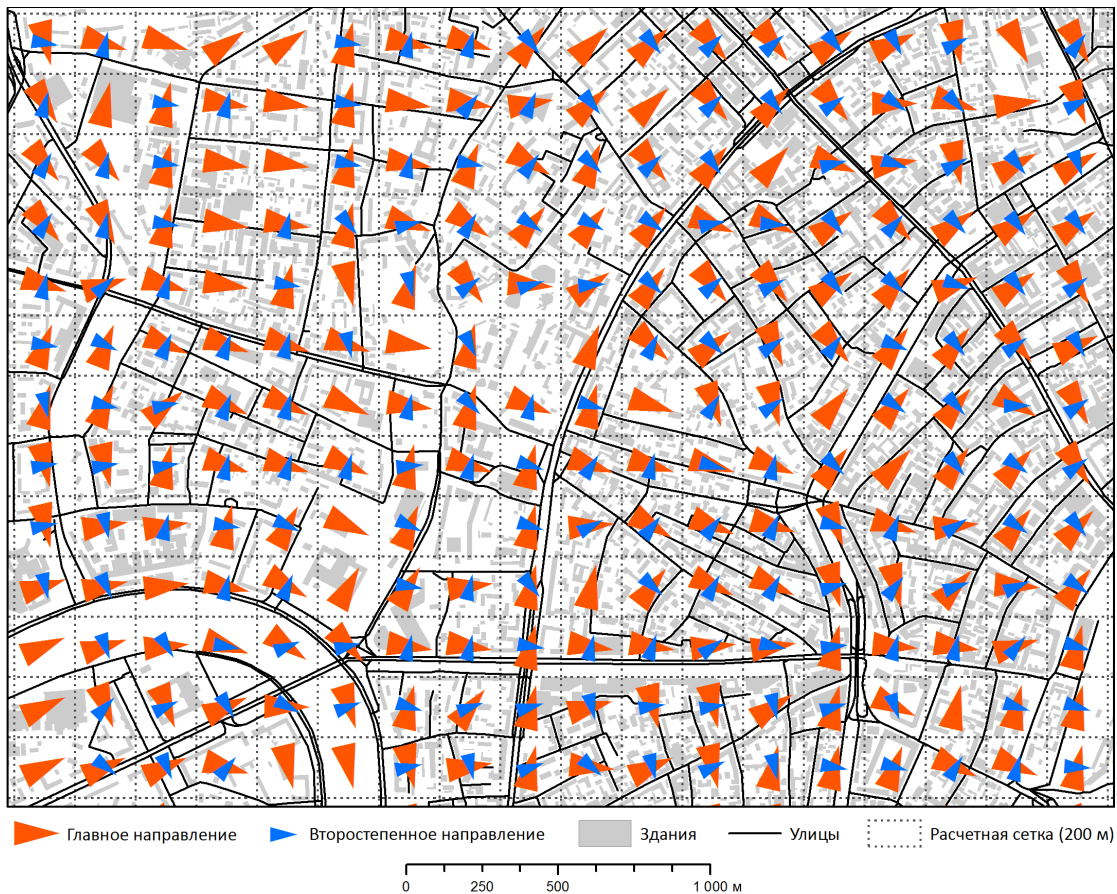

**Рис. 5. Преобладающие направления каньонов в ячейках**

## Параметры ячеек

К параметрам ячеек (36-42, Таблица 1) мы относим вспомогательную информацию, которая помогает специалистам метеорологами интерполировать данные более грубого разрешения (COSMO, WRF) на нашу расчетную сетку. Сюда входят прямоугольные, геодезические координаты центров ячеек, а также средняя абсолютная отметка ячейки (Рис. 6). Вычисление этих параметров выполняется стандартными средствами ГИС и не составляет труда. Поля SHAPE_Length и SHAPE_Area являются системными и формируются автоматически.

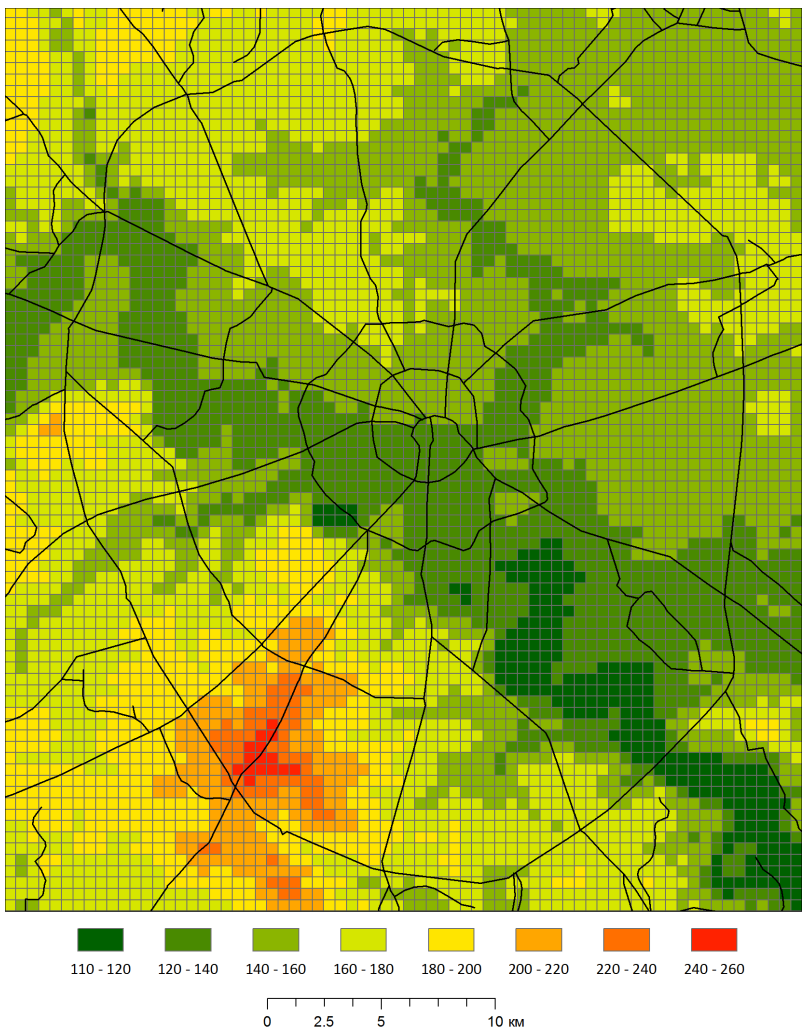

**Рис. 6. Средняя высота ячейки над уровнем моря (по данным SRTM). Шаг сетки 500 м.**

## Средние значения вычисленных характеристик по территории города Москвы и их чувствительность к разрешению

На основе созданной базы данных были подсчитаны статистические показатели по некоторым наиболее важным характеристикам (Таблица 2). Расчет производился по ячейкам, находящимся в пределах МКАД. В первую очередь, следует отметить стабильность распределения типов подстилающей поверхности при смене шага сетки. Колебания их площадей не превышают 1%. Среднестатистическая ячейка подстилающей поверхности в Москве 0,5х0,5 км состоит из водных объектов (3,4%), зданий (6,2%), дорог (12,8%), промышленных территорий (10%), озелененных территорий (18,9%) и неопределенных (прочих) категорий подстилающей поверхности (48,7%).

Существенно более чувствительны к разрешению характеристики геометрии городского каньона. Можно заметить, что увеличение шага сетки с 200 до 1000 м приводит к сжатию среднестатистического каньона (на 22% и 31% для направленных и ненаправленных каньонов) и снижению его доли внутри ячейки. Соответственно уменьшается и доля зданий, образующих направленные каньоны. В то же время, последовательная детализация масштаба моделирования как раз предполагает увеличение роли городских каньонов на более дробных сетках. Следовательно, полученный тренд средних значений отвечает целям работы. Если же говорить о конкретной средней ширине направленного и ненаправленного каньона, то она для Москвы составляет 60 и 40 метров соответственно при средней высоте здания 17 метров. Средняя высота зданий, формирующих направленные каньоны вдоль улиц по результатам расчетов равна 19.5 м, в то время как здания, не выходящие на улицы имеют среднюю высоту 13 м. Это можно объяснить значительной долей внутри кварталов невысоких сооружений, таких как школы, детские сады, трансформаторные подстанции и проч.

**Использование базы данных при метеорологическом моделировании**

Для включения созданной базы данных в качестве параметра метеорологической модели URB-MOS, она была экспортирована в набор текстовых файлов, отформатированных в соответствии с требованиями к входным данным. Таким образом, был определен интерфейс передачи данных. Уже первые эксперименты, проведенные с моделью URB-MOS (пока она остается единственной, из тех, кто может использовать созданную базу данных) показали улучшение качества воспроизведения модельных полей температуры и ветра. Сравнение результатов моделирования стандартной версии региональной модели Гидрометцентра РФ COSMO_RU с версией, включающей URB-MOS наблюдениями на московской метеостанции Балчуг (см. Рис. 7) в апреле 2011 года показало, что уточнённый с помощью модели URB-MOS прогноз значительно ближе к наблюдениям, чем исходный прогноз COSMO-RU, по обоим параметрам. Также стоит отметить, что среднеквадратичная ошибка прогноза температуры уменьшилась с 4.0 до 2.1 °C.

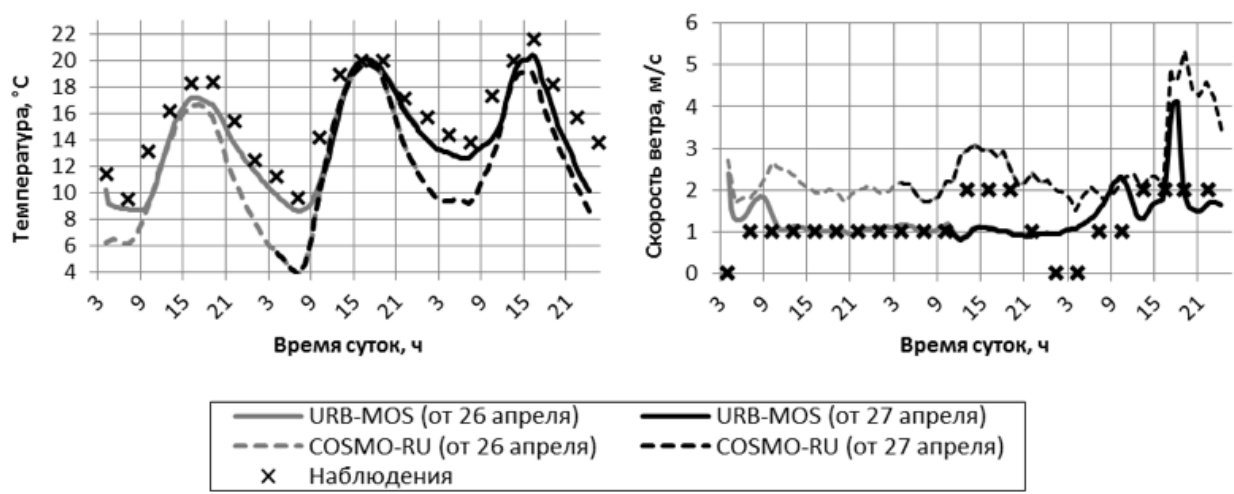

**Рис. 7. Сравнение исходного прогноза COSMO-RU, уточненного с помощью модели URB-MOS прогноза и натурных наблюдений для метеостанции Балчуг**

Таким образом, идея использования специализированной базы данных параметров подстилающей поверхности для уточнения и детализации прогноза погоды для города, а также для изучения эффекта городского острова тепла представляется весьма перспективной. Несмотря на множество недостатков модели URB-MOS, таких, как грубое описание водных и растительных ячеек и примитивные численные схемы, она успешно справилась с поставленной задачей. Такой результат, скорее всего, связан с более точным учетом как физических характеристик подстилающей поверхности (альбедо, теплоемкость), так и с прогрессивным методом моделирования термической адвекции и диффузии, который стал возможен только при использовании вышеописанной методики.

**Заключение**

При геоинформационных исследованиях подстилающей поверхности в городской среде используются в основном данные дистанционного зондирования. В то же время, целый ряд параметров, относящихся к геометрии городского каньона, требует привлечения картографических данных. В работе показано, каким образом с использованием методов пространственного анализа могут быть вычислены разнообразные характеристики подстилающей поверхности и городских каньонов.

1. Определен список параметров, необходимых для параметризации метеорологических моделей мегаполисов.
2. Разработана методика вычисления характеристик подстилающей поверхности и городского каньона с использованием методов пространственного анализа по картографическим данным.
3. Получена мультимасштабная база данных характеристик подстилающей поверхности и городского каньонов города Москвы с разрешением 200, 500 и 1000 м.
4. Определен интерфейс и принципы использования полученной БД при метеорологическом моделировании.
5. Произведена первичная оценка результатов метеорологического моделирования, полученных с использованием подготовленной БД

Выполненные нами расчеты опираются на несколько допущений. Принятые соотношения естественных и заасфальтированных поверхностей для промышленных территорий, а также характеристика озелененных территорий весьма условны и требуют уточнения по данным дистанционного зондирования, хотя и не вносят существенного вклада в погрешности результатов моделирования. Помимо этого принимается, что направленный каньон совпадает с линиями улиц. И хотя это в большинстве случаев именно так, могут быть потеряны направленные каньоны, образуемые зданиями, между которыми в картографических данных нет линий дорог.

Важнейшим результатом работы стала сформированная база данных характеристик подстилающей поверхности и городских каньонов Москвы,

которая будет использована в дальнейших экспериментах для параметризации моделей прогноза погоды и климатических моделей мегаполисов. Предложенные в работе методы вычисления характеристик универсальны и должны позволить сформировать аналогичные массивы данных по другим крупным городам России.

**Библиография:**